\documentclass[twocolumn,english,aps,prc,groupedaddress,amsmath,amssymb,showpacs]{revtex4}
\usepackage[T1]{fontenc}
\usepackage[latin9]{inputenc}

\makeatletter

\providecommand{\tabularnewline}{\\}

\@ifundefined{textcolor}{}
{%
 \definecolor{BLACK}{gray}{0}
 \definecolor{WHITE}{gray}{1}
 \definecolor{RED}{rgb}{1,0,0}
 \definecolor{GREEN}{rgb}{0,1,0}
 \definecolor{BLUE}{rgb}{0,0,1}
 \definecolor{CYAN}{cmyk}{1,0,0,0}
 \definecolor{MAGENTA}{cmyk}{0,1,0,0}
 \definecolor{YELLOW}{cmyk}{0,0,1,0}
 }

\makeatother

\usepackage{babel}
\begin{document}

\title{Clustering of energy levels}

\author{L. Zamick}

\author{A. Escuderos}

\affiliation{Department of Physics and Astronomy, Rutgers University, Piscataway,
New Jersey, 08854}
\begin{abstract}
It is noted that in single j-shell calculations certain odd-spin states
in even--even nuclei lie in a narrow energy band; likewise certain
states in odd--odd nuclei. 
\end{abstract}
\maketitle

\section{Results}

Our purpose here is to point out some striking results that occur
when single $j$-shell calculations are performed. Examples from the
$f_{7/2}$ and $g_{9/2}$ shells are given. We will not here make
comparison with experiment in part because many of the states in question
have not been found. But this should not stop us from trying to display
the simplicities that arise from complex calculations.

We first consider odd-spin states in $^{44}$Ti, $^{52}$Fe, and $^{96}$Cd.
We previously published results on these nuclei~\cite{ze12}. Here
we wish to highlight some interesting behaviour. We list in Table~\ref{tab:even}
the angular momenta and energies of odd-spin states that lie close
in energy. The input two-body matrix elements for $^{44}$Ti and $^{52}$Fe
were obtained from the spectra of $^{42}$Sc and $^{54}$Co respectively.
The two-body matrix elements of Corragio et al.~\cite{ccgi12} were
used for $^{96}$Cd.

\begin{table}[htb]
 \caption{\label{tab:even} Energies of close-lying states of odd spin in even--even
nuclei.}

\begin{ruledtabular} %
\begin{tabular}{ccc}
 & $J$  & $E$ (MeV) \tabularnewline
\hline 
$^{44}$Ti  &  & \tabularnewline
 & 1  & 5.669 \tabularnewline
 & 3  & 5.786 \tabularnewline
 & 5  & 5.871 \tabularnewline
 & 7  & 6.043 \tabularnewline
\hline 
$^{52}$Fe  &  & \tabularnewline
 & 1  & 5.518 \tabularnewline
 & 3  & 5.872 \tabularnewline
 & 5  & 6.271 \tabularnewline
 & 7  & 5.972 \tabularnewline
\hline 
$^{96}$Cd  &  & \tabularnewline
 & 1  & 4.269 \tabularnewline
 & 3  & 4.467 \tabularnewline
 & 5  & 4.566 \tabularnewline
 & 7  & 4.635 \tabularnewline
 & 9  & 4.565 \tabularnewline
\end{tabular}\end{ruledtabular} 
\end{table}

We do not have any obvious explanation of this simple behaviour. There
has not been too much focus on odd-$J$ even-parity states in even--even
nuclei.There is no experimental information on these odd-spin states.
We hope that our highlighting this simple behaviour will stimulate
further research on this topic.

As a second example, we consider a system of five neutrons and one
proton ($^{46}$Sc) and of one neutron-hole and five proton-holes
($^{94}$Rh). The energies of select states are shown in Table~\ref{tab:odd}.
The same two-body matrix elements that were used in Table~\ref{tab:even}
were used here.

\begin{table}[htb]
 \caption{\label{tab:odd} Energies of close-lying states of odd--odd nuclei.}

\begin{ruledtabular} %
\begin{tabular}{ccc}
 & $J$  & $E$ (MeV) \tabularnewline
\hline 
$^{46}$Sc  &  & \tabularnewline
 & 2  & 0.0000 \tabularnewline
 & 3  & 0.0987 \tabularnewline
 & 4  & 0.0664 \tabularnewline
 & 5  & 0.2805 \tabularnewline
 & 6  & 0.0355 \tabularnewline
\hline 
$^{94}$Rh  &  & \tabularnewline
 & 2  & 0.5207 \tabularnewline
 & 3  & 0.5285 \tabularnewline
 & 4  & 0.3282 \tabularnewline
 & 5  & 0.5215 \tabularnewline
 & 6  & 0.2033 \tabularnewline
 & 7  & 0.5253 \tabularnewline
 & 8  & 0.0000 \tabularnewline
 & 9  & 0.7183 \tabularnewline
\end{tabular}\end{ruledtabular} 
\end{table}

We see for $^{46}$Sc four states are calculated to be below 100~keV
whilst for $^{94}$Rh four states near 0.52~MeV. Quite remarkable.
Again we do not have a great explanation for this behaviour, but feel
that it is worth pointing this out. 

In $^{46}$Sc the experimental values from J=2+ to J=6+ are respectively0.4441,
0.2278, 0.0000, 0.2807 and 0.0520 MeV. Although agreement with experiment
is far from perfect the feature of clustering is clealy present. In
$^{94}$Rh the angular momenta with known energies are J=2+, 4+ ,8+
and 9+. The respective energies in MeV are 0.0546, 0.0000, x+0.0 and
x+ 0.5675. Here also the agreement with experiment is not satisfactory
but the feature of clustering is present. This clustering phenomenum
should become a focus of future studies.

\end{document}